\documentclass[conference]{IEEEtran}
\IEEEoverridecommandlockouts
\usepackage{cite}
\usepackage{amsmath,amssymb,amsfonts}
\usepackage{algorithmic}
\usepackage{graphicx}
\usepackage{textcomp}
\usepackage{xcolor}

\def\BibTeX{{\rm B\kern-.05em{\sc i\kern-.025em b}\kern-.08em
    T\kern-.1667em\lower.7ex\hbox{E}\kern-.125emX}}

\usepackage{amsthm}
\usepackage{balance}
\usepackage{dblfloatfix}

\newtheorem{theorem}{Theorem}

\begin{document}

\title{Quantum Circuit Pruning: Improving Fidelity via Compilation-Aware Circuit Approximation \vspace{-10pt}
\thanks{
This work was supported by the European Commission (QUADRATURE: 101099697, WINC: 101042080), the Spanish Ministry of Science, Innovation and Universities and the European ERDF (PID2021-123627OB-C51, PID2024-158682OB-C31, QCOMM-CAT), the QuantERA project EQUIP (PCI2022-133004) funded by the Gobierno de España (MCIN/AEI/10.13039/501100011033), MICIN and NextGenerationEU (PRTR-C17.I1), the Generalitat de Catalunya, the ICREA Academia Award 2024, and the predoctoral FPI-UPC grant (UPC–Banco Santander).
}
}

\author{
    \IEEEauthorblockN{
        Pau Escofet\IEEEauthorrefmark{1},\
        Santiago Rodrigo\IEEEauthorrefmark{1},\
        Rohit Sarma Sarkar\IEEEauthorrefmark{1},\\
        Carmen G. Almudéver\IEEEauthorrefmark{2},\
        Eduard Alarcón\IEEEauthorrefmark{1},\ and
        Sergi Abadal\IEEEauthorrefmark{1}
    }
    \IEEEauthorblockA{
        \textit{\IEEEauthorrefmark{1}Universitat Politècnica de Catalunya, Spain} \qquad\quad
        \textit{\IEEEauthorrefmark{2}Universitat Politècnica de València, Spain}\\
        pau.escofet@upc.edu
        \vspace{-15pt}
    }
}

\maketitle

\begin{abstract}
This work presents a routing-aware pruning strategy for quantum circuits executed on Noisy Intermediate-Scale Quantum (NISQ) devices. We propose a method to remove parametric controlled rotations whose small rotation angles do not justify the routing overhead required for their implementation. By selectively pruning such gates, the method mitigates fidelity loss arising from additional \texttt{SWAP} operations introduced during compilation. Our approach evaluates whether executing a gate leads to greater fidelity loss than omitting it. Simulations on benchmark circuits with realistic noise models show that the method reduces two-qubit gate counts (up to 48.6\%) while improving final state fidelity (up to 47.7\%), especially for larger circuits where routing costs dominate.
\end{abstract}


\section{Introduction}
Quantum computing holds the promise of solving problems that are intractable for classical computers, with potential breakthroughs in areas such as cryptography, optimization, chemistry, and machine learning \cite{nielsen_chuang_2010, montanaro2016quantum}. While large-scale fault-tolerant quantum computers remain a long-term goal, significant progress is being made with Noisy Intermediate-Scale Quantum (NISQ) devices, systems containing tens to hundreds of noisy qubits that are not yet capable of integrating full-fledged quantum error correction protocols \cite{preskill_quantum_2018}. The reliability of computations on NISQ devices is severely constrained by noise and limited qubit connectivity. To maximize the utility of current hardware, a vast body of research has focused on improving overall circuit's fidelity through advances in noise mitigation and in compilation, including qubit mapping and routing, and circuit optimization \cite{murali2020software, tannu2019not, li_2019_tackling}.

In this work, we introduce a qubit routing-aware pruning strategy that evaluates each parametric two-qubit gate in terms of its expected impact on the quantum state relative to the routing cost required for its execution. We assess the proposed method on a suite of benchmark circuits compiled to grid-based architectures under a realistic noise model. Simulations show that our approach can remove up to 25–40\% of two-qubit gates while improving final circuit fidelity in most cases, with gains becoming more pronounced as circuit size and routing complexity increase. These results demonstrate that routing-aware pruning can be a practical and scalable approach to enhancing computation reliability on NISQ hardware.

\section{Background}
Current quantum processors suffer from limited qubit connectivity, imperfect gate operations, and significant decoherence, all contributing to reducing quantum state fidelity, and therefore, impacting their computational capability \cite{preskill_quantum_2018, arute2019quantum}.

Efficiently executing quantum circuits on connectivity-constrained quantum devices requires a qubit mapping process in which logical qubits (from the circuit) are mapped to physical qubits (from the processor), often inserting \texttt{SWAP} gates to enable interactions between non-adjacent qubits \cite{qubit_allocation}. While routing is necessary for executing circuits, it results in a non-negligible two-qubit gate overhead that can significantly degrade performance (\textit{i.e.} circuit fidelity) \cite{murali2019noise}, especially in deep circuits.


To mitigate these effects, various pruning strategies have been proposed, including techniques that eliminate gates with negligible overall effect on the quantum state \cite{barenco1996approximate, imamura2023offline, kulshrestha2024qadaprune}. Existing pruning approaches are typically carried out in the pre-routing stage and consider only the rotation angle of the parametric gate, ignoring the actual connectivity constraints and routing overhead. This gap motivates the development of our routing-aware pruning strategy, which integrates pruning decisions directly into the compilation flow.

Such pruning can be particularly effective when targeting gates whose implementation requires substantial routing effort. More precisely, a parametric controlled rotation gate with a negligible angle of rotation applied to neighboring qubits can be executed without additional \texttt{SWAP} gates (“free” in terms of routing cost). Therefore, its removal does not contribute to improving the overall circuit's fidelity.

In this work, we conjecture that integrating pruning decisions into the routing process, specifically targeting small-angle parametric controlled rotations that incur high routing overhead, can improve circuit fidelity on NISQ devices compared to pre-routing pruning strategies.

\section{Methods}
In this section we devise a methodology for pruning gates from a quantum circuit based on the extra operations (\textit{e.g.} \texttt{SWAPs}) required for routing the qubits to neigbouring positions. We aim to maximize the overall circuit fidelity, and compare the expected fidelity loss from imperfect routing operations, to the loss incurred by omission of the parametric controlled rotations.

We define the fidelity of a $\theta$-rotation over the $\hat{n}$ axis ($F_{R_{\hat{n}}(\theta)}$) as the maximum change over all quantum states $|\psi\rangle$, thus leading to a minimum fidelity between $|\psi\rangle$ and $R_{\hat{n}}(\theta) |\psi\rangle$:
\begin{equation}
    F_{R_{\hat{n}}(\theta)} := \min_{|\psi\rangle} F(|\psi\rangle, R_{\hat{n}}(\theta) |\psi\rangle)
    \label{eq:f_op}
\end{equation}
where $F(|\psi\rangle, R_{\hat{n}}(\theta)|\psi\rangle)$ is the quantum fidelity \cite{nielsen_chuang_2010} between the original and the rotated stated, computed as:
\begin{equation}
    F(|\psi\rangle, R_{\hat{n}}(\theta)|\psi\rangle) = \left|\langle\psi|R_{\hat{n}}(\theta)|\psi\rangle \right|^2
\end{equation}
On the other hand, the cost in fidelity of routing quantum states across the quantum computer by inserting \texttt{SWAP} gates depends on the error rate of each quantum gate and the distance (\textit{i.e.} number of \texttt{SWAP} gates that need to be added) between the physical qubits to be routed ($d_{q_i,q_j}$).
\begin{equation}
    F_{\texttt{swap}} := \left( (1-p_2)^{3 \cdot \left\lceil \frac{d_{q_i, q_j}}{2} \right\rceil} + \frac{1 - (1-p_2)^{3 \cdot \left\lceil \frac{d_{q_i, q_j}}{2} \right\rceil}}{4} \right)^2
    \label{eq:f_swap}
\end{equation}
The above expression (following \cite{Escofet_2025}) assumes that both qubits $q_i$ and $q_j$ have been moved towards the middle position between them, effectively applying $\lceil{d_{q_i, q_j}/2}\rceil$ \texttt{SWAP} gates ($3\cdot \lceil{d_{q_i, q_j}/2}\rceil$~\texttt{CNOTs}) to each qubit, each gate adding depolarizing noise with depolarization parameter $p_2$. Since the employed routing strategy is not optimal (routing is an NP-complete problem \cite{qubit_allocation}), the actual number of \texttt{SWAP} gates inserted typically exceeds $d_{q_i,q_j}$. To account for this, we approximate the number of \texttt{SWAP} gates required for an interaction between qubits $q_i$ and $q_j$ to be $1.25 \cdot d_{q_i,q_j}$. Equation~(\ref{eq:f_swap}) showcases the fidelity of a composite quantum state after applying $3 \cdot\lceil{d_{q_i, q_j}/2}\rceil$ gates to each qubit, following the model for fidelity under depolarizing noise proposed in \cite{Escofet_2025}.

\subsection{Parametric Gates}
A parametric gate is a quantum gate whose operation depends on a continuous parameter, such as a rotation angle, phase, or amplitude. These gates allow fine control over quantum state evolution, often used to represent variable-strength interactions in quantum algorithms, and are fundamental building blocks for VQAs \cite{cerezo2021variational} and QML \cite{biamonte2017quantum}. Common examples include single-qubit rotations $R_X(\theta)$ or $R_Z(\theta)$, as well as controlled-rotation gates like $CR_Z(\theta)$ or $CP(\theta)$. In this work, we focus on parametric controlled rotation gates where the rotation angle $\theta$ determines the magnitude of the transformation, making them suitable candidates for pruning when $\theta$ is small relative to the routing overhead.

With $\hat{n} = (n_x, n_y, n_z)$, a single qubit rotation by $\theta$ about the $\hat{n}$ axis can be defined by \cite{nielsen_chuang_2010}:
\begin{align}
    R_{\hat{n}} (\theta) &= e^{-i \frac{\theta}{2} \hat{n} \overrightarrow{\sigma}}\\
    &= \cos\left(\frac{\theta}{2}\right) I -i \sin\left(\frac{\theta}{2}\right)\underbrace{\left( n_x X + n_y Y + n_z Z \right)}_{R_{\hat{n}}}
\end{align}
where $\overrightarrow{\sigma}$ denotes the three component vector ($X, Y, Z$).

Now, for an arbitrary rotation $R_{\hat{n}}(\theta)$ over a particular axis $\hat{n}$ applied to $|\psi\rangle$, we want to quantify how much does the state changes depending on $\theta$.

\begin{theorem}
Given a $\theta$-rotation to an arbitrary quantum state $|\psi\rangle$, the fidelity between the rotated and unrotated state is lower-bounded by:
\begin{equation}
     F(|\psi\rangle, R_{\hat{n}}(\theta)|\psi\rangle) \geq \cos^2 \left( \frac{\theta}{2} \right)
     \label{eq:theo_op}
\end{equation}
\label{theo:rotation}
\end{theorem}

\begin{proof}
    \vspace{-20px}
    \begin{align}
         F(|\psi\rangle, &R_{\hat{n}}(\theta)|\psi\rangle)\\
          &= \left|\langle\psi|R_{\hat{n}}(\theta)|\psi\rangle \right|^2 \\
          &= \left| \langle\psi| \left[ \cos\left(\frac{\theta}{2}\right) I -i \sin\left(\frac{\theta}{2}\right)R_{\hat{n}} \right] |\psi\rangle \right|^2\\
          &= \left| \cos\left(\frac{\theta}{2}\right) \langle\psi|\psi\rangle -i \sin\left(\frac{\theta}{2}\right) \cdot \langle\psi|R_{\hat{n}} |\psi\rangle \right|^2\\
          &= \cos^2\left(\frac{\theta}{2}\right) \langle\psi|\psi\rangle^2 +\sin^2\left(\frac{\theta}{2}\right) \cdot  \langle\psi|R_{\hat{n}} |\psi\rangle^2\\
          &\geq \cos^2\left(\frac{\theta}{2}\right) \cdot 1 +\sin^2\left(\frac{\theta}{2}\right) \cdot 0\\
          &= \cos^2 \left( \frac{\theta}{2} \right)
    \end{align}
\end{proof}

This lower bound on the fidelity between both states is sharp since we consider the case where $\langle\psi|R_{\hat{n}} |\psi\rangle = 0$, which in reality will depend on the exact state $|\psi\rangle$. Take $R_{\hat{n}} = X$ as an example (\textit{i.e.}, a rotation on the $x$-axis), then $F(|0\rangle, X|0\rangle) = 0$, while $F(|+\rangle, X|+\rangle) = 1$. Nevertheless, we know that the fidelity between two states cannot be negative, and therefore, by considering the case where $\langle\psi|R_{\hat{n}} |\psi\rangle = 0$, we explore the worst-case scenario regarding the quantum states difference.

\subsection{Routing-Aware Pruning}
Instead of eliminating (pruning) gates from the circuit based solely on the rotation angle they apply, such as methods like \cite{10.5555/984720.984755, barenco1996approximate, imamura2023offline, kulshrestha2024qadaprune}, in this work, we propose a routing-based pruning strategy based both on the rotation angle of the gate and its expected routing cost.

\begin{figure}
    \vspace{-10px}
    \centering
    \includegraphics[width=\linewidth]{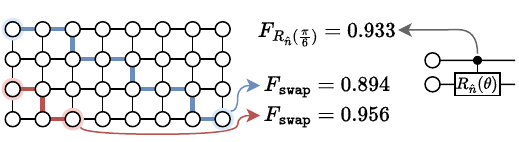}
    \vspace{-20px}
    \caption{Before applying a conditional rotation $\theta$, the compiler assesses the distance between the two interacting qubits (blue and red paths). The probability of pruning increases with path length. The values shown correspond to a $p_2 = 0.005$ ($F_{\texttt{swap}}$ from Equation~(\ref{eq:f_swap})) and a $\theta = \pi/6$ rotation ($F_{R_{\hat{n}}(\theta)}$ from Equation~(\ref{eq:theo_op})). In this example, a $\frac{\pi}{6}$-rotation between the blue qubits is pruned, while the same rotation between the red qubits is not.}
    \label{fig:path}
    \vspace{-15px}
\end{figure}

\begin{figure*}
    \centering
    \includegraphics[width=1\linewidth]{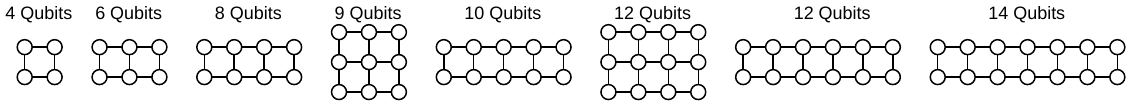}
    \vspace{-20px}
    \caption{Considered nearest-neighbor topologies ranging from 4 to 14 qubits.}
    \label{fig:systems}
    \vspace{-15px}
\end{figure*}

To this purpose, we compare $F_{R_{\hat{n}}(\theta)}$ and $F_{\texttt{swap}}$, defined in Equations (\ref{eq:f_op}) and (\ref{eq:f_swap}) respectively. Every time a parametric two-qubit gate needs to be executed, $F_{R_{\hat{n}}(\theta)}$ is calculated based on the $\theta$-rotation applied by the considered gate, and $F_{\texttt{swap}}$ is computed based on the current physical positions of the two qubits involved. If $F_{\texttt{swap}} < F_{R_{\hat{n}}(\theta)}$, the rotation introduced by the gate is not sufficiently significant to justify the routing overhead, and the gate is therefore discarded (see Figure \ref{fig:path}).

Circuits with parametric gates containing between 4 and 14 qubits are considered, obtained from the MQT Bench \cite{quetschlich2023mqtbench}. Circuits are routed using the Route-Forcing algorithm \cite{10821308} due to its link error rate awareness when routing, and its fast execution time. Regardless, the proposed pruning strategy can be incorporated to most qubit routing algorithms \cite{lao_2022_timing, 8342181, li_2019_tackling, zou2024lightsabre}. Circuits are routed to comply with the connectivity constraints of a grid topology \cite{google2025quantum} (\textit{i.e.} nearest-neighbor connectivity among physical qubits), as shown in Figure \ref{fig:systems}.

After routing (and pruning), the circuit is decomposed into the basis gate set, to match the platform feasible operations. In this work we use IBM's basis set gate \texttt{\{CX, ID, RZ, SX, X\}}. Note that two-qubit parametric gates must be pruned before the final basis gate decomposition, since in the used gate set there are no parametric two-qubit gates, which are decomposed into a combination of parametric single-qubit gates and \texttt{CX}.

\subsection{Simulation and Noise Model}
To fully assess the impact of the proposed pruning methodology under realistic conditions, we employ Qiskit \cite{Qiskit} to simulate the selected circuits, which highly limits the scale of the systems we can evaluate due to the complexity of quantum simulation on classical machines \cite{10.1145/3310273.3323053, 10.5555/3135595.3135617, Boixo2018}.

For each circuit, three states are computed. The \emph{Ideal State} is obtained from a noiseless, exact simulation of the circuit. The \emph{Noisy State} is considered to be a realistic noisy state for current architectures. It is obtained through the simulation of the compiled circuit under a realistic noise model, and it is used as a baseline. Finally, the \emph{Pruned State} is generated using the same compilation and noisy simulation procedure as the \emph{Noisy State}, but with the proposed pruning strategy applied to parametric gates. 

We then assess the fidelity between the \emph{Ideal} and \emph{Noisy States}, and compare it to the fidelity between the \emph{Ideal} and \emph{Pruned States}. If the latter is greater than the former, it means that, by pruning some gates in the circuit, we are able to alleviate the compilation overhead of the circuit and therefore reach a more accurate result.


Regarding the noise model, each two-qubit gate introduces depolarizing noise (with parameter $p_2$) and thermal relaxation effects (characterized by $T_1$ and $T_2$) on the involved qubits. Since the goal is to evaluate the effect of the proposed pruning strategy rather than reproduce the exact behavior of existing hardware, we select $p_2$, $T_1$, and $T_2$ values that result in a representative fidelity for the \emph{Noisy State}. The \emph{Pruned State} is then compared against this baseline to quantify any improvement using the same noise parameters.


For the following results, $p_2$ is set to
\begin{equation}
    p_2 =\frac{1}{ \left( \texttt{\#gates} / \texttt{\#qubits} \right)^{2}}
    \label{eq:p2}
\end{equation} 
and the thermal relaxation parameters are set to 
\begin{equation}
    T_1 = T_2 = 2 \cdot \texttt{circuit duration [s]}
\end{equation}
where circuit duration is defined as the cumulative time for executing all the gates in the circuit.

\section{Results}
Seven quantum circuits ranging from 4 to 14 qubits are simulated with the previously defined settings. For each circuit, we assess how does the pruning methodology proposed in this works affects the circuit, tracking how many \texttt{CX} gates are removed and how this pruning affects the final fidelity against the \emph{Ideal State}.

\begin{figure*}
    \centering
    \includegraphics[width=0.93\linewidth]{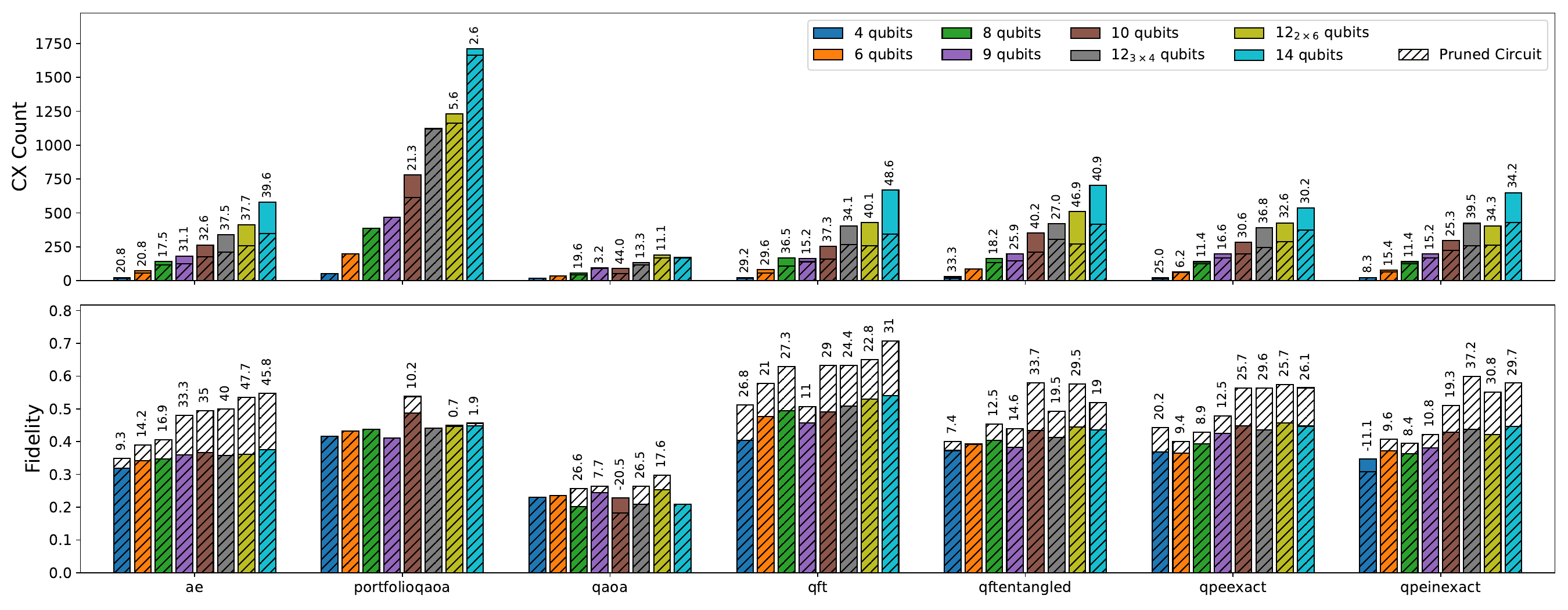}
    \vspace{-10px}
    \caption{Two-qubit gate count (top row) and circuit fidelity (bottom row) for the selected benchmarks containing between 4 and 14 qubits. The percentage improvement in fidelity achieved by the pruning methodology is indicated above each corresponding bar.}
    \label{fig:bar_fidelity}
    \vspace{-15px}
\end{figure*}

Figure \ref{fig:bar_fidelity} showcases the number of \texttt{CX} gates and fidelity against the \emph{Ideal State} (noiseless) for the circuits after compilation, when using our pruning technique. It can be seen how, for most circuits, the proposed pruning methodology reduces the number of two-qubit gates after compilation (up to 48\%) while improving the circuit's final fidelity (up to 47\%), thus validating the initial hypothesis.


It is interesting to see the different behavior of the circuits. For example, pruning the \textit{Amplitude Estimation} (ae) algorithm, or the \textit{Quantum Fourier Transform} (\texttt{qft} / \texttt{qftentangled}), results in an increase in fidelity across all circuit sizes. On the other hand, for circuits like the \textit{Quantum Approximate Optimization Algorithm} (\texttt{qaoa} / \texttt{portfolioqaoa}) pruning only has an impact on particular sizes. We attribute this to an uneven connectivity between the different system sizes. Consider, for instance, the 10- and 12-qubit cases, implemented on $2\times5$ and $3\times4$ grids, respectively. Although the 12-qubit system includes more qubits, and thus more gates, leading to a lower effective $p_2$ (see Equation~(\ref{eq:p2})), both architectures share the same connectivity graph diameter.

\begin{figure}
    \centering
    \includegraphics[width=0.93\linewidth]{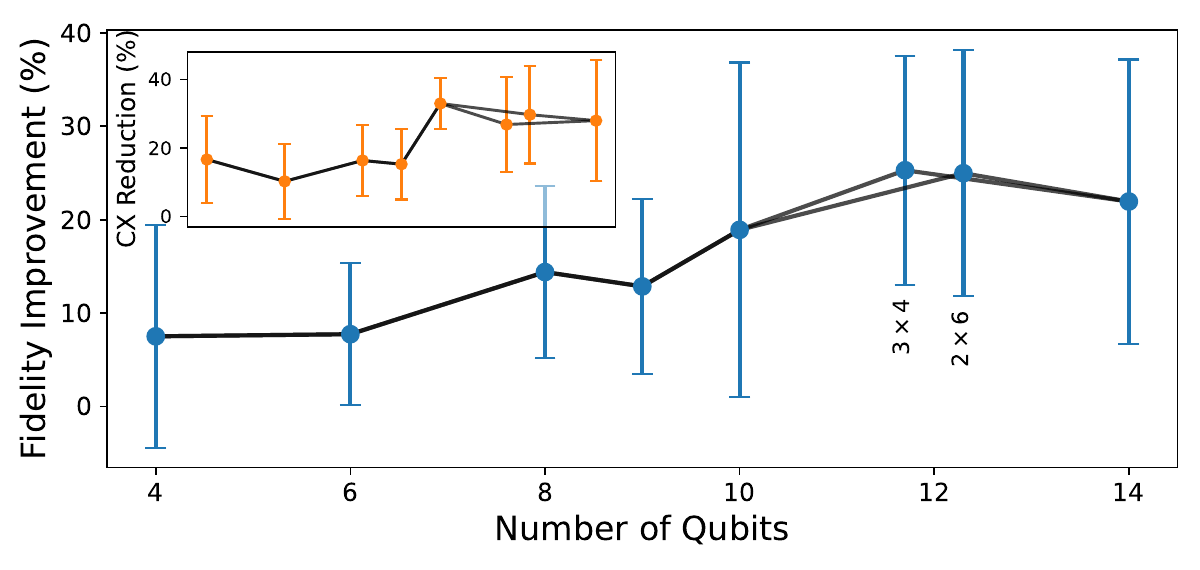}
    \vspace{-10px}
    \caption{Fidelity's relative improvement when pruning the circuit (blue) and relative reduction of two-qubit gates (orange) for an increasing circuit size.}
    \label{fig:size_fidelity}
    \vspace{-15px}
\end{figure}

To assess the improvement that pruning brings, Figure \ref{fig:size_fidelity} shows, for each system size, the relative improvement in fidelity (\%), reporting mean and standard deviation. These results show a higher fidelity enhancement as circuit size grows. This trend arises from the increasing difficulty of routing as the system size grows. Longer paths lead to more gates being pruned, suggesting that the proposed method will become increasingly relevant for larger quantum systems, those required to achieve quantum advantage \cite{preskill_quantum_2018}.

Lastly, we analyze the performance of our proposed methodology against a compilation-agnostic pruning for the QFT circuit characterized by an approximation degree, proposed by Qiskit \cite{Qiskit} and based on \cite{barenco1996approximate, 10.5555/984720.984755}. In such approximations the gates with the \texttt{approximation\_degree} smallest rotation angles are removed from the circuit.

\begin{figure}
    \centering
    \includegraphics[width=0.93\linewidth]{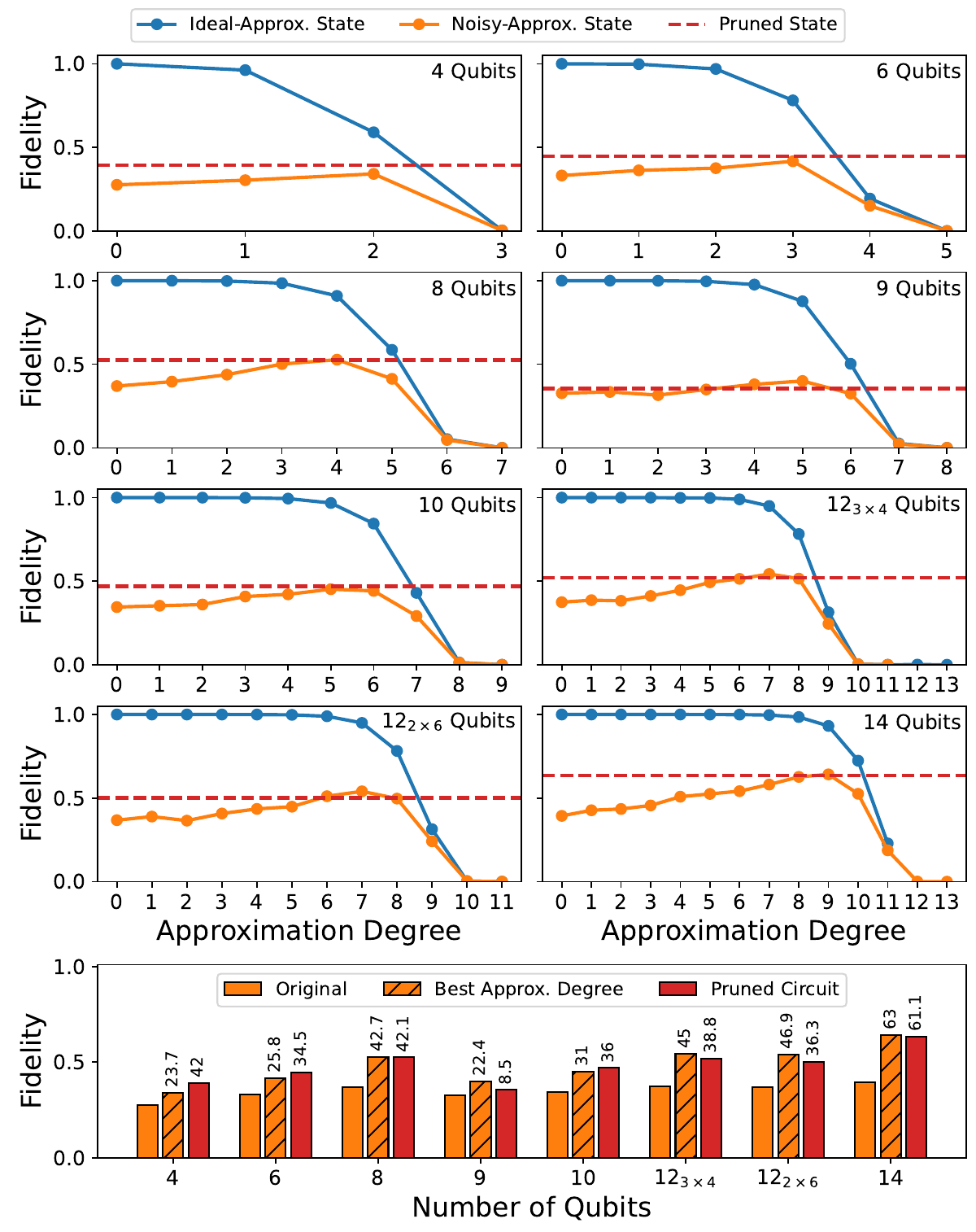}
    \vspace{-10px}
    \caption{Fidelities for the noiseless and noisy approximation (blue and orange lines, respectively), and this work's proposed pruned circuit (red), for QFT circuit sizes between 4 and 14 qubits. The percentage improvement in fidelity for the best approx. degree and the pruned circuit is indicated above each corresponding bar.}
    \label{fig:qft}
    \vspace{-12px}
\end{figure}

The results depicted in Figure \ref{fig:qft} show that the methodology proposed in this work matches (in terms of fidelity) the best approximation degree for most system sizes. It is worth noting that the best approximation degree depends on the error parameters, the processor topology, and other factors, needing of experimental exploration before knowing which one performs the best, an unfeasible task for larger systems, which are, by nature, not simulable in classical machines.

Overall, the results reveal that incorporating routing costs into pruning decisions provides consistent benefits across different algorithms and system sizes, without requiring prior tuning of approximation parameters. The proposed approach adapts naturally to the circuit structure and hardware connectivity, delivering near-optimal fidelity improvements in scenarios where traditional pruning either removes beneficial “free” gates or fails to eliminate costly long-range interactions. This adaptability makes it a promising strategy for future compilers targeting larger and noisier quantum devices.

\section{Conclusions}
We have introduced a routing-aware pruning methodology that selectively removes two-qubit parametric gates whose expected contribution to the computation is outweighed by the fidelity loss from routing overhead. Simulations demonstrate that this approach consistently reduces two-qubit gate counts (up to 48.6\%) and can improve final state fidelity (up to 47.7\%), particularly for larger circuits where routing overhead is significant. These results highlight the potential of integrating routing-aware pruning into quantum compilers to improve the reliability of computations on NISQ hardware. Future work will extend this method and evaluate its performance on real quantum processors.

\newpage
\bibliographystyle{IEEEtran}
\balance
\bibliography{IEEEabrv, main}

\end{document}